\documentclass{osa-article}

\journal{josab}

\articletype{Research Article}

\begin{document}

\title{Broadband lasers for photo-ionization and repumping of trapped ions}

\author{T. Fordell\authormark{*} and T. Lindvall}

\address{VTT Technical Research Centre of Finland Ltd, Centre for Metrology MIKES, P.O. Box 1000, FI-02044 VTT, Finland}

\email{\authormark{*}thomas.fordell@vtt.fi} 



\begin{abstract}
A frequency-stable, broadband laser is presented for experiments on trapped ions. Since the design is based on widely available semiconductor optical amplifier technology, similar lasers can be realized for virtually any wavelength in the near-infrared, and the coherence properties and output power allow for efficient second harmonic generation. No closed-loop frequency stabilization for addressing Doppler- or naturally-broadened, dipole-allowed transitions is needed, and the light source can be turned on and off during a measurement cycle with sub-microsecond response time. As a case study, a 921.7-nm laser with an output power of 20~mW and a linewidth of 10~GHz is realized, which is then frequency doubled to 460.9-nm for excitation of strontium as the first step in photo-ionization. The excitation efficiency is compared to that achievable with a narrow-linewidth distributed Bragg reflector laser as well as to theory.
\end{abstract}

\section{Introduction}
Trapped ions are used in various fields of physics, e.g., in quantum information processing \cite{Haffner2008a, Wineland2011a} mass spectrometry \cite{March2009, Nolting2017} and optical time and frequency metrology \cite{Margolis2009a, Poli2013a, Ludlow2015b}. Producing the ions and manipulating their states require a considerable ensemble of lasers, electro-optics and control electronics. Especially considering transportable experiments and future more wide spread use of these technologies, it is of considerable interest to find simple and robust approaches for each part of the experimental setups.

 There is a large number of experiments on ions without hyperfine structure,  $^{40}$Ca$^+$, $^{88}$Sr$^+$ and $^{138}$Ba$^+$ being the most common ones. In addition to lasers for photo-ionization, most of these experiments incorporate lasers for the $^2D_{3/2} \rightarrow {^2}P_{1/2}$ transition (at 866~nm, 1092~nm, and 650~nm, respectively) and for the ${^2}D_{5/2} \rightarrow {^2}P_{3/2}$ transition (at 854~nm, 1033~nm, and 614~nm, respectively). Some also use the $^2D_{3/2} \rightarrow {^2}P_{3/2}$ transition (at 850~nm, 1004~nm, and 586~nm, respectively).  The lack of hyperfine structure means that the light sources employed do not need to be narrowband in order to distinguish states within a narrow hyperfine manifold. 
 
 In \cite{Lindvall2013a, Fordell2015a} the use of turn-key, broadband light sources based on amplified spontaneous emission were proposed and demonstrated for repumping and clock state clear out for a $^{88}$Sr$^{+}$ single-ion clock. These sources require no frequency stabilization and can be electronically turned on and off during the measurement cycle; however, while the presented schemes work well for the transitions in question, the achievable power spectral densities (PSD) remain low and insufficient for weaker transitions. Here, simple schemes for achieving PSDs and modulation bandwidths orders of magnitude higher are presented. As a case study, the 460.9-nm
$^1 S_0 \rightarrow {^1}P_1$ transition in neutral strontium is driven, which is the first step in producing ions via two-step photo-ionization \cite{Kjaergaard2000a, Gulde2001a}. In this first step, atoms are excited by a laser, which can provide isotope selectivity (see e.g. \cite{Johanning2011a,Kjaergaard2000a,Lucas2004a,Tanaka2005a,Wang2011b}). For Sr, however, the frequency shift between the $^1 S_0 \rightarrow {^1}P_1$ transition in $^{88}$Sr (natural abundance 83\%) and the nearest hyperfine component (corresponding to $^1P_1,\; F=7/2$) in the third most abundant isotope $^{87}$Sr (7\%) is only $-8$~MHz~\cite{Bushaw2000a}. Considering the natural linewidth $\Delta\nu = 30.24$\;MHz \cite{Yasuda2006a} of the transition, this makes fully isotope selective loading of $^{88}$Sr from a natural mixture difficult, but fortunately its abundance is high. All the less abundant isotopes ($^{86}$Sr, $^{87}$Sr, $^{84}$Sr) can be selectively photo-ionized using a narrow-linewidth laser orthogonal to a well-collimated atomic beam and with a sufficiently low intensity to avoid saturation broadening. An isotopically enriched source would, of course, largely eliminate loading problems relating to isotopes.

In the second photo-ionization step, an extremely broad ($\sim 1$\;nm) transition involving an auto-ionizing state can be used, which can be driven with cheap, unstabilized (blue-ray) laser diodes \cite{Brownnutt2007a}. In the case of Ca \cite{Schuck2010a} and Ba \cite{Wang2011b}, light-emitting diodes (LEDs) have also proved effective.

The laser architecture presented below together with the ASE light sources \cite{Lindvall2013a, Fordell2015a} means that all 'auxiliary' light sources (excitation, ionization, repumper and clear out) needed for operating a Sr$^+$ clock can be built into a small form factor without extensive and expensive customization and operated intermittently in a clock cycle without active frequency stabilization. The general design principles in this work were to construct a light source in the near infrared (921.7~nm) with a spectral line broad enough to accommodate the frequency drift remaining after simple passive frequency stabilization but still narrow enough to enable efficient second harmonic generation (SHG) to the blue part of the spectrum (460.9~nm) and to provide sufficient PSD.

\section{Setup}

\begin{figure}[t]
\includegraphics[width=0.45\linewidth]{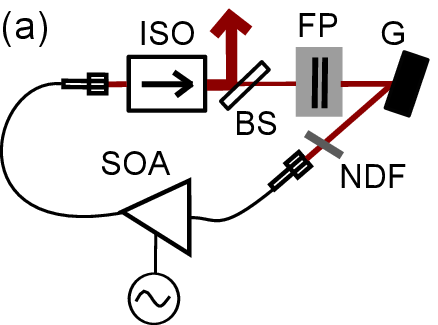}
\hspace{0.05\linewidth}
\includegraphics[width=0.45\linewidth]{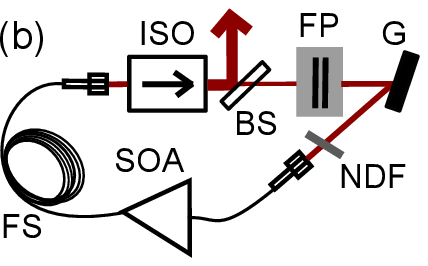}
\caption{Broadband ring laser with (a) and without (b) modulation. ISO: optical isolator; BS: beam splitter; FP: temperature-controlled Fabry-P\'erot etalon; G: grating; SOA: semiconductor optical amplifier (Superlum SOA-472); NDF: neutral-density filter; FS: fiber spool.}
\label{fig:setup}
\end{figure}

\begin{figure}[t]
\includegraphics[width=.48\linewidth]{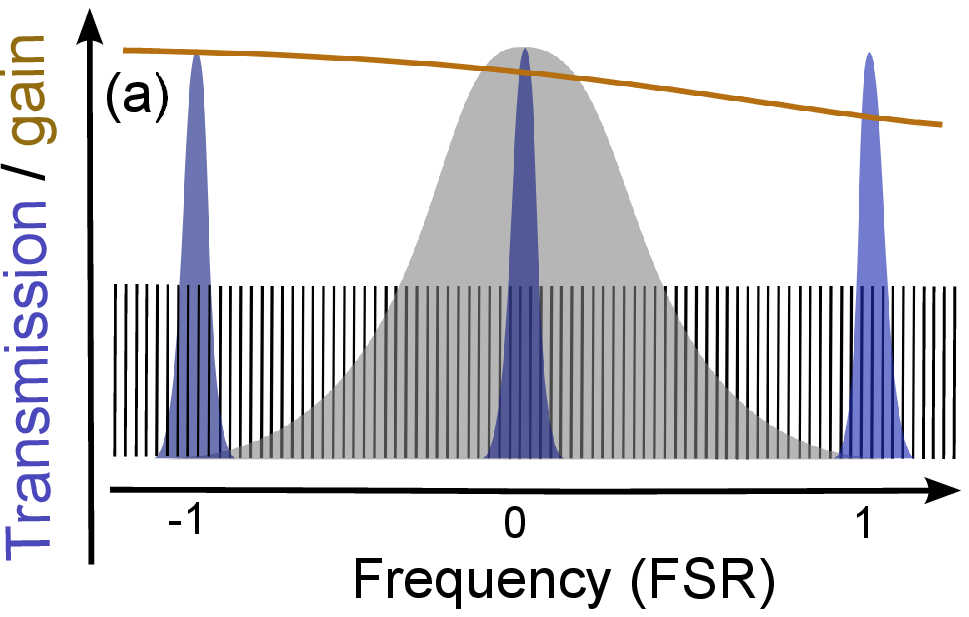}
\includegraphics[width=.48\linewidth]{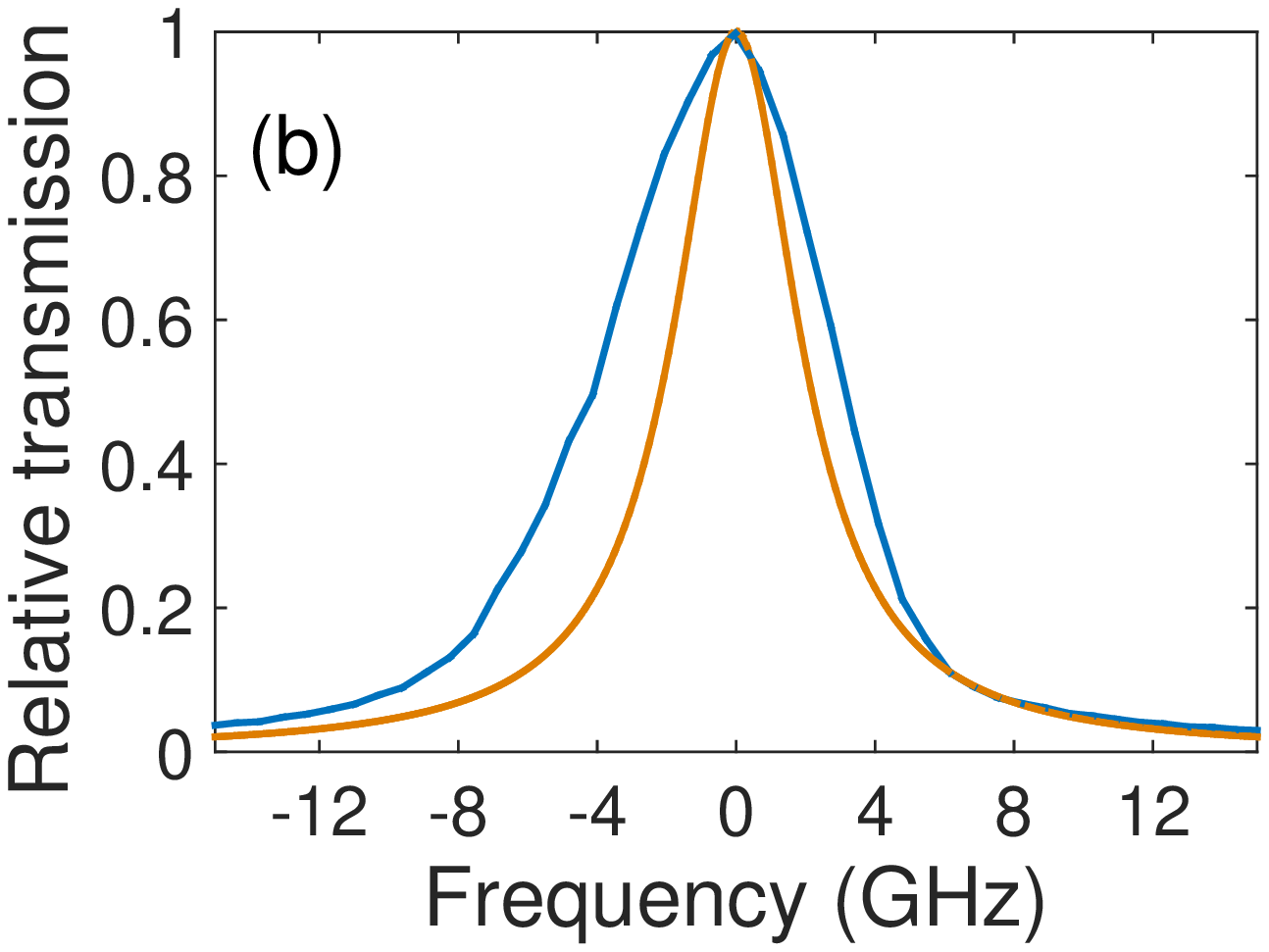}
\caption{(a): Schematic of spectral filtering. A ruled diffraction grating provides coarse filtering (gray) of the longitudinal cavity modes (black) under the SOA gain envelope (brown). Centre frequency and bandwidth is determined by a temperature-controlled thin solid-glass etalon (blue). (b) Experimental  (blue) and theoretically expected (brown) transmission for the 0.5-mm fused silica etalon used in this work.}
\label{fig:filtering}
\end{figure}

Achieving high PSD with low total power output means that the spectral bandwidth has to be relatively small; thus, very narrowband spectral filtering and subsequent amplification of high-power super-luminescent LEDs that have recently become widely available is not an option, since the microwatts of power available after filtering will not be sufficient for properly seeding an amplifier. Since adding filtering and amplification stages increases cost and complexity, what is needed is a high PSD to begin with, which is only available through lasing action. While several different cavity configurations can be used in constructing a laser, here we have implemented a ring cavity (Fig.~\ref{fig:setup}) with a semiconductor optical amplifier (SOA) as the gain element. For low input levels, a SOA in a ring cavity behaves as an inhomogeneously broadened gain medium, resulting in multimode operation with a mode spacing determined by the cavity length. Control of the bandwidth and center wavelength is achieved in two steps (Fig.~\ref{fig:filtering}a): a grating for coarse tuning at the nanometer level, and a simple, temperature-controlled planar etalon for fine tuning and wavelength stabilization. Here the etalon consists simply of a 0.5-mm thick disc of fused silica (parallelism <10") coated on both sides with a 94\% reflective coating. The experimental and theoretical transmission of this etalon at normal incidence is shown in Fig.~\ref{fig:filtering}b. The measured finesse is about 30, surprisingly close to the theoretical value of 48 considering the planar-planar geometry.

The center frequency of the etalon is sensitive to temperature, pressure and the angle of incidence. Ignoring the thin coatings, the temperature sensitivity is calculated to be 2~GHz/K, pressure sensitivity as small as 3~kHz/Pa, and the sensitivity to the angle of incidence 50~GHz/deg; therefore, it is possible to control the frequency drift to $\ll 1$~GHz by simply controlling the temperature of the small glass plate and by ensuring that the input coupling is mechanically stable. This, in turn, means that if the linewidth is, say, 5-10~GHz, then no frequency stabilization other than a modest control of the etalon temperature is needed. In this work, the etalon is suspended between two silicone o-rings and the temperature of the housing is controlled using a highly-stable glass-bead thermistor as the temperature sensor \cite{Lawton2001a, Lawton2002a}.

To be able to reliably address dipole-allowed transitions with a natural linewidth of a few tens of MHz, the gaps of 250~MHz between the longitudinal modes need to be removed. There are two approaches to this: adding noise or modulation into the bias current of the amplifier (Fig.~\ref{fig:setup}a) or increasing the cavity length (Fig.~\ref{fig:setup}b). A noisy current source would be the method of choice for a short cavity; however, when using discrete components with fibre coupling as in this work, the ratio of cavity length to gain medium length becomes so large that the modulation would need to be extremely strong; therefore, a 10-m fibre spool was added in order to reduce the mode spacing to 19~MHz (Fig.~\ref{fig:setup}b). Approximately 90\% of the power available after the isolator was coupled out of the cavity, which turned out to be too small a fraction. A neutral density filter (OD 0.5) had to be inserted before the amplifier; otherwise occasional jumps to single mode operation were observed when optimizing the cavity alignment.

\section{Results}

\begin{figure}[t]
\includegraphics[width=0.5\linewidth]{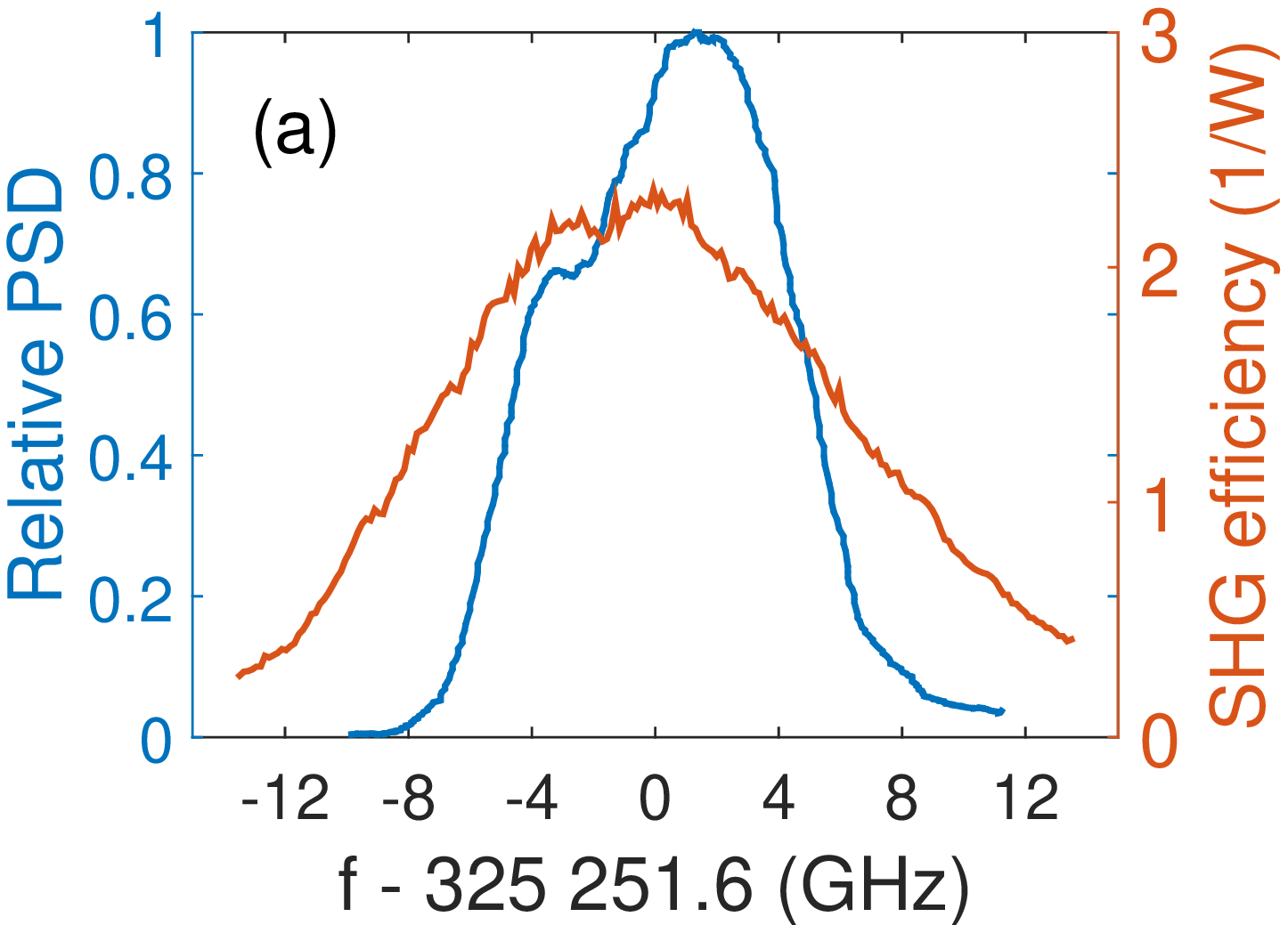}
\includegraphics[width=0.5\linewidth]{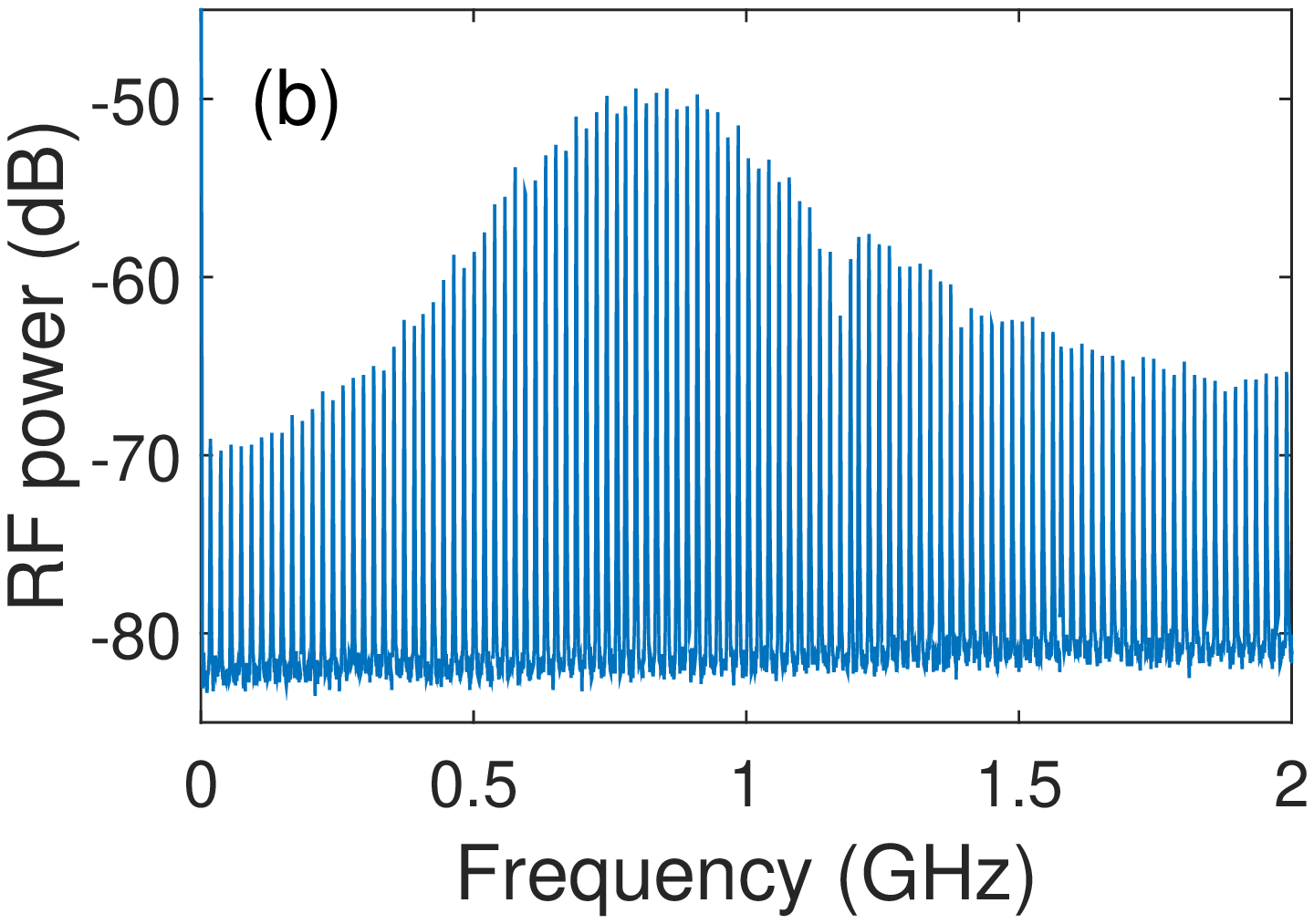}
\caption{(a) Optical spectrum of the broadband laser (blue) and SHG efficiency of the periodically-poled lithium niobate waveguide (red). (b) RF spectrum showing the beat note between hundreds of longitudinal modes.}
\label{fig:spectrum}
\end{figure}

The output beam quality is very good owing to the single-mode-fiber-coupled SOA. The spectrum at an output power level of 20~mW is shown in Fig.~\ref{fig:spectrum}a (blue trace), which was obtained by scanning a single-mode distributed Bragg reflector (DBR) laser across the spectrum while recording the beat note using a radio frequency (RF) spectrum analyzer and the DBR laser wavelength using a wavemeter. The RF power was averaged over one longitudinal mode spacing (19~MHz) in order to bring out the envelope and avoid clutter due to the drifting longitudinal modes underneath it. The mode-averaged PSD is 3-4 orders of magnitude higher than in the un-amplified broadband sources presented in \cite{Fordell2015a}. The comb-like RF power spectrum, shown in Fig.~\ref{fig:spectrum}b, reveals the beating between the hundreds of longitudinal modes.

While the center frequency of the broadband laser (BBL) is mainly determined by the etalon, the spectral width is also affected by the SOA gain, i.e., by its operating temperature and bias current, and by losses in the rest of the cavity. Close to threshold, the sensitivity of the spectral width to these parameters will of course be large, but when the laser is driven well above threshold, the changes become minor. Given an etalon, the spectral width can be adjusted by several GHz by selecting a suitable combination of bias current and cavity loss. The higher the SOA gain, the higher the output power and the larger the spectral width. In this work, the SOA was operated close to maximum bias current in order to have efficient SHG. 

Since the spectral properties are largely determined by the etalon and since multimode operation is allowed, 
 the SOA does not require a stable, low-noise current controller. Indeed, the spectrum in Fig. \ref{fig:spectrum}a does not change even when the SOA is powered directly from a laboratory power supply (voltage source) with only a resistor limiting the current. In the cavity configuration used throughout this work, the length of the gain element is but a small fraction of the cavity length (Fig. \ref{fig:setup}b), which means that the behaviour of the individual longitudinal modes underneath the spectral envelope is also largely unaffected; however, in a short-cavity configuration, this would become close to what is depicted in Fig. \ref{fig:setup}a, where a noisy current source is used to broaden the modes to close the spectral gaps between them. Also in stark contrast to alternative lasers based on semiconductor gain media, such as DBR and distributed feedback (DFB) lasers, the gain element does not require a temperature controller (TEC). Disengaging the TEC of the SOA only results in reduced gain and, therefore, in lowered power output and in minor changes to the spectrum as explained above, the centre frequency still being determined by the etalon. A corollary of the insensitivity to the gain media temperature is that the BBL can be turned on and off by modulating the SOA bias current with sub-microsecond response times. Contrary to this, the DBR laser used above to characterize the spectrum takes several tens of seconds to settle ones the bias current is turned on, even if the TEC is engaged continuously. This temperature insensitivity means that in non-power-critical applications, the TEC can also be dropped. In this work, maximum power output was required for efficient SHG, so a TEC was used to keep the gain element slightly below room temperature.

\begin{figure}[t]
\includegraphics[width=0.5\linewidth]{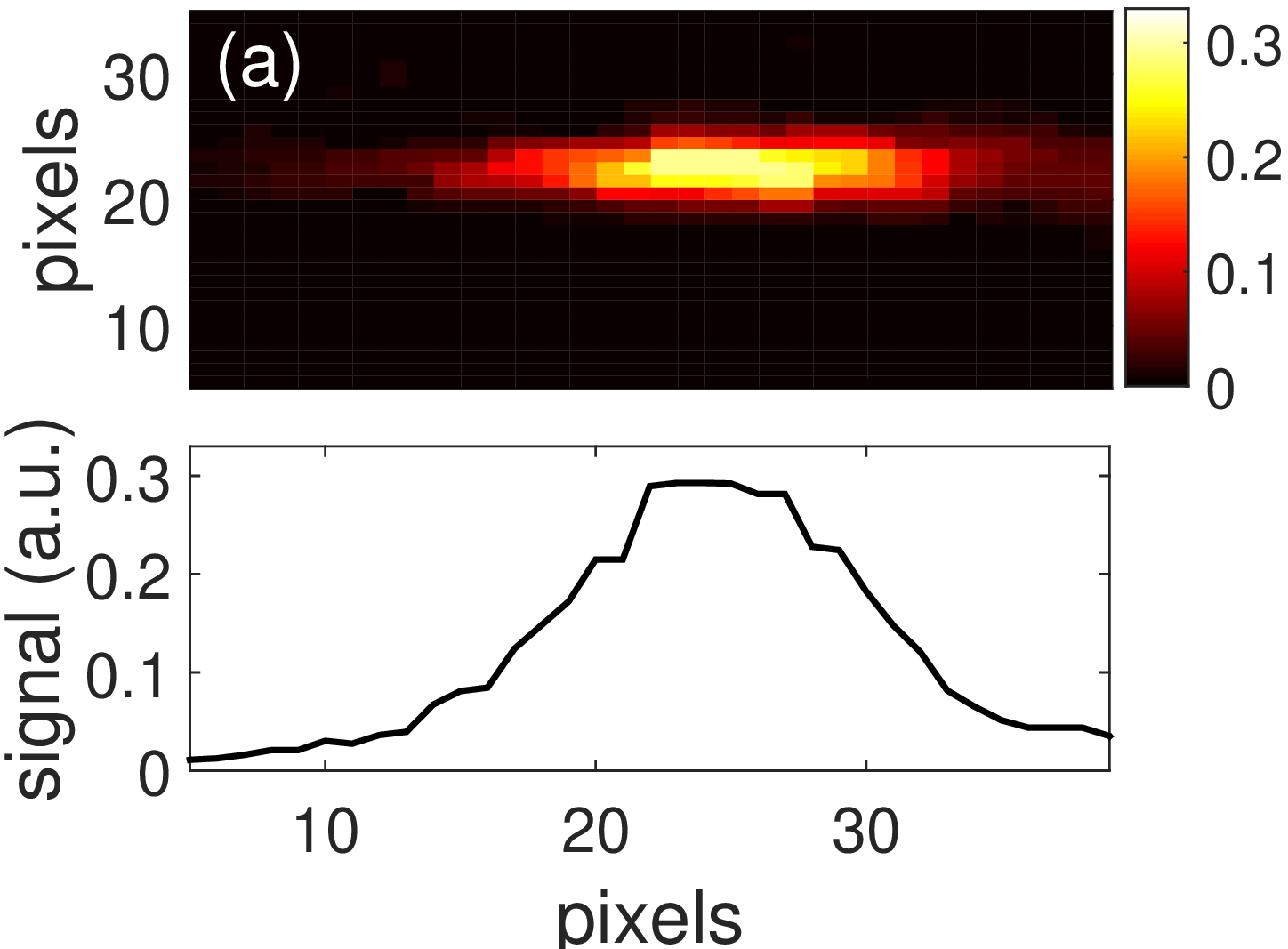}
\includegraphics[width=0.5\linewidth]{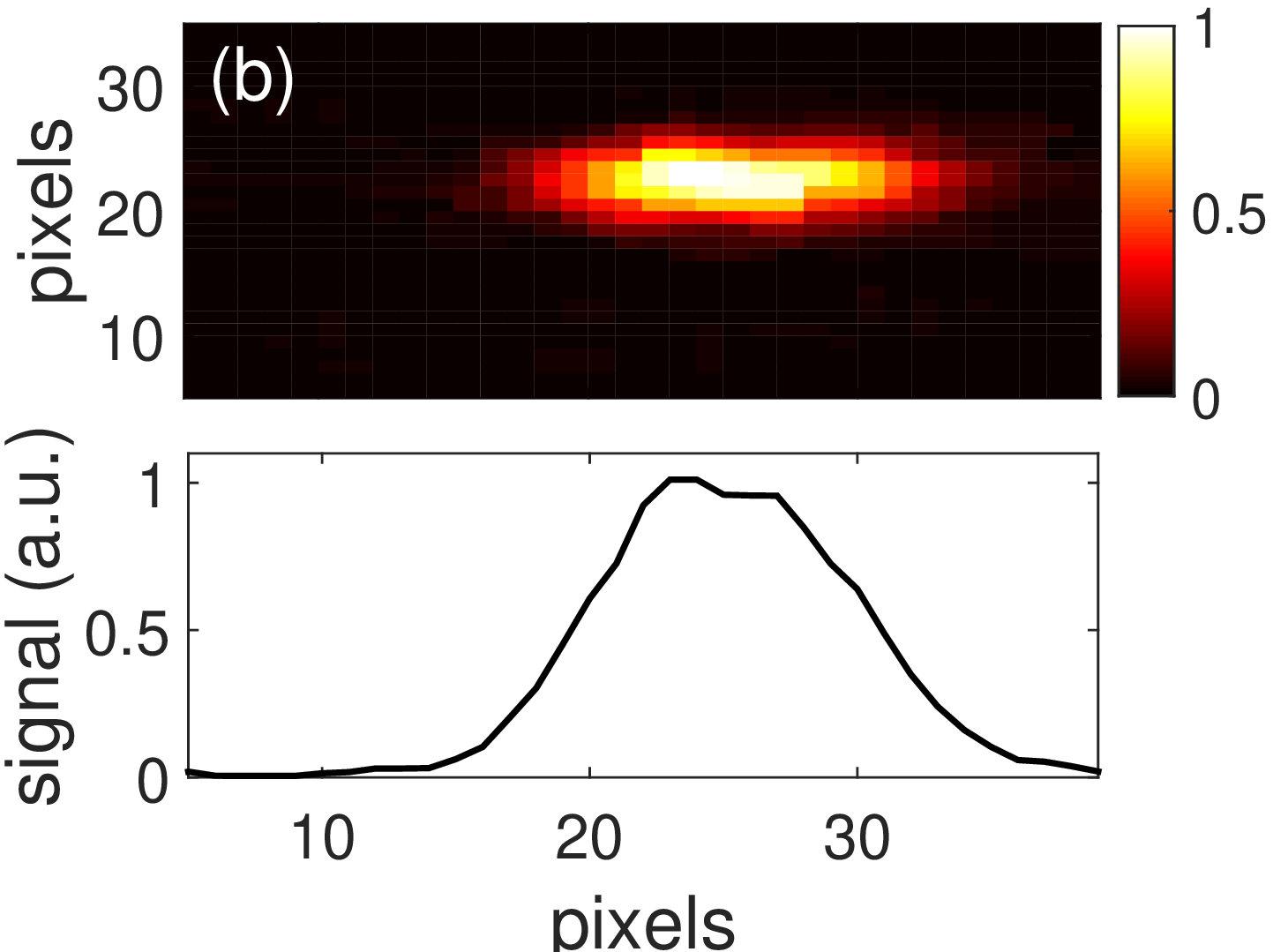}
\caption{Strontium fluorescence due to the broadband source (a) and the DBR laser (b) as seen by a CCD camera with a $\sim$15-degree viewing angle with respect to the laser beam. In (b), the laser was tuned to maximum efficiency. Note the factor of three difference in colour scales. The bottom graphs show cross-sections through maximum fluorescence. The width of the images correspond to ${\sim}7$\;mm. }
\label{fig:flourescence}
\end{figure}

In order to check the excitation probability of the $^1 S_0 \rightarrow {^1}P_1$ transition in Sr, the near infra-red light was frequency doubled in a 22-mm-long periodically-poled lithium niobate waveguide (NTT Electronics WH-0461-000-A-B-C) to 460.9~nm. The SHG efficiency is shown in Fig.~\ref{fig:spectrum}a (red trace) for the DBR laser. Since the coherence length of the broadband source is still larger than the optical length of the doubling crystal, only a small ($\sim$20\%) drop in efficiency was observed compared to the DBR laser. The frequency doubled light was sent at near-normal incidence into a thermal beam of Sr produced by a Sr dispenser (Alvatec AS-2-Sr-45-F) inside a small vacuum cell, and the fluorescence was recorded by a camera positioned behind the exit window. The angle between the optical axis of the camera objective and the laser beam was only $\sim$15\;degrees; still, the fluorescence due to the broadband source (Fig.~\ref{fig:flourescence}a) appears clearly elongated compared to that of the laser (Fig.~\ref{fig:flourescence}b), which indicates that a much wider part of the velocity distribution is being addressed. At an equal power level of 1~mW (0.88\;mW inside vacuum cell) and with a beam waist $w = 170$\;\textmu m, the excitation efficiency due to the DBR laser is estimated to be approximately 3.3 times stronger. According to the theory presented below, the efficiencies would be almost equal for a beam waist of 40\;\textmu m.

\section{Theoretical fluorescence rates}

In the following, we derive simple equations comparing the excitation probability (or fluorescence rate) of a BBL compared to that of a narrow-band laser for interaction with an ensemble of Doppler-broadened two-level atoms. These relations can be used to estimate the amount of BBL power needed to replace a narrow-band laser in a certain application.

The numerical results below correspond to the experiment described above. The $^1 S_0 \rightarrow {^1}P_1$ transition in $^{88}$Sr is well approximated by a two-level system,
as the branching ratio to the $^1$D$_2$ level is very small ($\sim 1:5\times 10^4$) \cite{Dinneen1999a} and the transition amplitudes to all three Zeeman sublevels are the same $C = 3^{-1/2}$ (for a definition of $C$, see Eq.~(7) in \cite{Lindvall2012a}). The nearly transverse ($\sim$86\;degrees) velocity distribution of the Sr flux from the dispenser was measured using the DBR laser and was found to be a Maxwell-Boltzmann distribution,
\begin{equation}
p_v(v) = \frac{1}{\sqrt{\pi} v_\mathrm{mp}} \exp{\left(-\frac{v^2}{v_\mathrm{mp}^2}\right)},
\end{equation}
with a FWHM Doppler width of $\Delta\omega_\mathrm{D} = 2 (\ln{2})^{1/2} k v_\mathrm{mp} = 2\pi\times 1.4$\;GHz, where $k$ is the wave number. This corresponds to a most probable velocity of $v_\mathrm{mp} = (2k_\mathrm{B}T/m)^{1/2} = 390\;\mathrm{m}/\mathrm{s}$, where $k_\mathrm{B}$ is the Boltzmann constant and $m$ the atomic mass. This is equal to a temperature $T = 800$\;K (in agreement with the dispenser datasheet).

The characteristic interaction time of the atoms passing through the laser beam is $\sim 2w/(\sqrt{2} v_\mathrm{mp}) = 600$\;ns ($\sqrt{2} v_\mathrm{mp}$ is the longitudinal rms velocity of the atomic beam). As this is much longer than the timescale at which the internal state of the atoms reaches steady state, $\Gamma^{-1} = (2\pi\Delta\nu)^{-1} = 5$\;ns, we can use the steady-state solution of the density matrix equations. The excited-state population of a two-level atom with the velocity $v$ along the laser beam direction is then given by
\begin{equation}\label{eq:rho_ee}
\rho_{ee}(v) = \frac{(\Gamma^*/\Gamma)(\Omega/2)^2}{(\delta-kv)^2+(\Gamma^*/2)^2+2(\Gamma^*/\Gamma)(\Omega/2)^2}.
\end{equation}
Here $\Omega$ is the Rabi frequency and $\delta$ is the detuning from resonance for an atom with zero velocity along the laser beam direction. 
The effect of a finite laser linewidth (spectrum assumed to be Lorentzian for simplicity) is accounted for by setting the dephasing rate of the optical coherence to $\Gamma^*/2$, where $\Gamma^* = \Gamma + \Delta\omega_\mathrm{laser}$ is the sum of the natural linewidth and the linewidth of the laser \cite{Blushs2004a}. If the linewidth of the laser is much narrower than the natural linewidth, we obtain the usual two-level expression.

For the DBR laser, we have $\Delta\omega_\mathrm{DBR} \ll \Gamma \ll \Delta\omega_\mathrm{D}$. Thus, when calculating the velocity-averaged population $\langle \rho_{ee}\rangle = \int_{-\infty}^{\infty} \rho_{ee}(v) p_v(v) dv$, the velocity distribution will be approximately constant over the range around $v = \delta/k$ where the Lorentzian is non-zero and it can be taken out of the integral. Integrating over the Lorentzian then gives
\begin{equation} \label{eq:rho_DBR}
\langle \rho_{ee}\rangle_\mathrm{DBR} = \frac{\sqrt{\pi}}{k v_\mathrm{mp}} \frac{(\Omega/2)^2}{(\Gamma/2)\sqrt{1+2(\Omega/\Gamma)^2}},
\end{equation}
where we have assumed to be near resonance, $\delta \ll k v_\mathrm{mp}$, and have set $\Gamma^* = \Gamma$.

For the broadband laser, on the other hand, we have $\Gamma \ll \Delta\omega_\mathrm{D} \ll \Delta\omega_\mathrm{BBL}$. This means that the Doppler shift in the denominator of Eq.~(\ref{eq:rho_ee}) is negligible and no integration over the velocity distribution is required. If we additionally assume $\delta \ll \Delta\omega_\mathrm{BBL}$, the excited-state population becomes
\begin{equation} \label{eq:rho_BBL}
\rho_{ee,\mathrm{BBL}} = \frac{\Omega^2}{\Delta\omega_\mathrm{BBL}\Gamma \left[1+2\Omega^2/(\Gamma\Delta\omega_\mathrm{BBL})\right]},
\end{equation}
i.e., it depends only on $\Omega^2/\Delta\omega_\mathrm{BBL}$, which is proportional to the peak PSD of the (assumed) Lorentzian spectrum, $\mathrm{PSD}_0 = 2P/(\pi\Delta\omega_\mathrm{BBL})$, where $P$ is the total power. We therefore choose $\Delta\omega_\mathrm{BBL}/2\pi = 11$\;GHz to make $\mathrm{PSD}_0$ equal to the actual experimental PSD at resonance. 


The Rabi frequency can be evaluated as $\Omega = C \frac{\mu}{\hbar} \sqrt{\frac{2I}{\epsilon_0 c}}$, where $\hbar$ is the reduced Planck constant, $\epsilon_0$ is the vacuum permittivity, and $c$ is the speed of light \cite{Lindvall2012a} . The reduced matrix element is $\mu = 5.248 e a_0$ \cite{Yasuda2006a,Safronova2013a}, where $e$ is the elementary charge and $a_0$ is the Bohr radius. We neglect the Gaussian beam profile and use the average intensity $I = P/(\pi w^2)$. For the waist $w = 170$\;\textmu m, this gives the Rabi frequency in units of $\Gamma$ as $\Omega/\Gamma = \sqrt{P/73\;\mu\mathrm{W}}$.

\begin{figure}[t]
\includegraphics[width=1\linewidth]{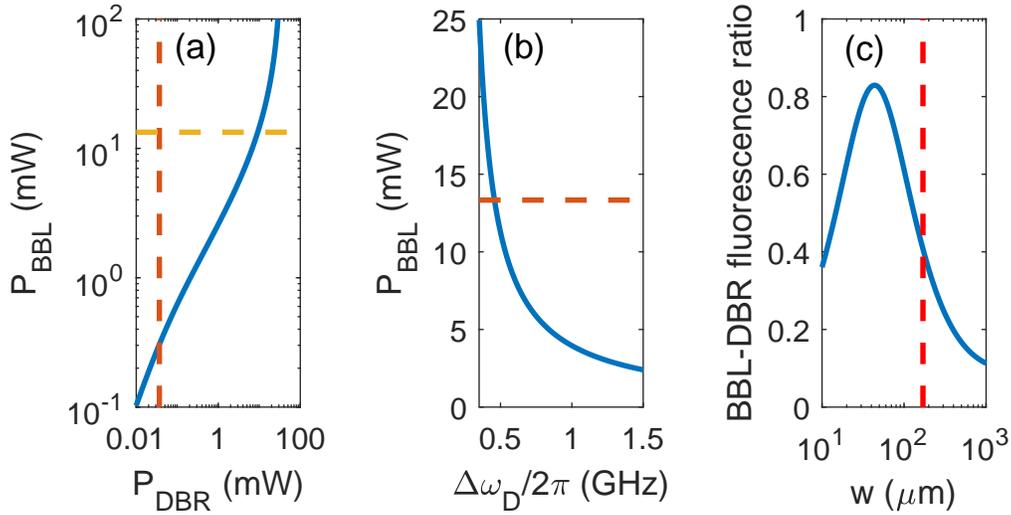}
\caption{(a) BBL power required to give same velocity-averaged fluorescence rate as DBR laser power on horizontal axis for the velocity distribution and waist size in the experiment (solid). The vertical and horizontal dashed lines show the saturation powers of the DBR laser and the BBL, respectively. (b) BBL power required to give same velocity-averaged fluorescence rate as 1\;mW of DBR laser power. The dashed line is the saturation power. (c) Expected BBL to DBR laser fluorescence ratio as a function of beam waist. The dashed red line indicates the experimental beam waist.}
\label{fig:theory}
\end{figure}

Using Eqs.~(\ref{eq:rho_DBR}--\ref{eq:rho_BBL}), we can then calculate the ratio of the fluorescence rates for the DBR laser and the BBL, which for $P = 0.88$\;mW becomes $2.5$, in reasonable agreement with the experimental value of 3.3 considering the simple models used and the experimental uncertainties. Equating Eqs.~(\ref{eq:rho_DBR}) and (\ref{eq:rho_BBL}), we can solve the BBL power required to obtain the same velocity-averaged fluorescence rate as for a certain DBR power. Figure~\ref{fig:theory}(a) shows this for the Doppler width and beam waist used in the experiment. We see that the use of a BBL is the most efficient in the intermediate range, where the DBR is affected by saturation, but the BBL is not. Figure~\ref{fig:theory}(b) shows the BBL power required to obtain the same velocity-averaged fluorescence rate as for 1\;mW of DBR power as a function of Doppler width over the range where both Eq.~(\ref{eq:rho_DBR}) and Eq.~(\ref{eq:rho_BBL}) are valid. Finally, Fig.~\ref{fig:theory}(c) shows how the BBL-DBR fluorescence ratio $\langle \rho_{ee} \rangle_\mathrm{BBL}/\langle \rho_{ee}\rangle_\mathrm{DBR}$ depends on the beam waist.

If one wants to use a BBL with cold ions or atoms, Eq.~(\ref{eq:rho_BBL}) is still valid. The comparison with a narrowband laser can in this case be made using Eq.~(\ref{eq:rho_ee}) by setting $v=0$. With the narrowband laser at resonance ($\delta = 0$), the BBL power required for the same excited-state population (or fluorescence rate) is then simply given by $\Delta\omega_\mathrm{BBL}/\Gamma$ times the power of the narrowband laser. To saturate a dipole-allowed transition in a cold ion a power level of the order of 1~\textmu W is typically needed, so for a BBL with a 10-GHz linewidth, the corresponding value is of the order of 1~mW.

\section{Discussion}

Set up on an optical breadboard without an enclosure and with discrete, free-space components and spring-loaded optics mounts, the proof-of-principle broadband laser has been tested intermittently for almost a year in a laboratory environment. During this time the spectral properties and power output have turned out to be very robust. The observed frequency drift (<2\;GHz) has been well below the linewidth, and most of this drift, if not all, is believed to be caused by changes in beam alignment as a result of testing different ND filters, moving the SOA fiber pigtails, and due to vibrations in connection with the spring-loaded mounts. 

Especially for transportable experiments, robustness can be further improved by replacing the bulk optics with fibre optics, that is, with a fibre Bragg grating (FBG) replacing the grating, a fused splitter for output coupling as well as a fiber-coupled isolator. The grating could also be replaced with a narrowband interference filter or sensitivity to its orientation could be reduced by using shorter etalons, since this increases their free-spectral range (FSR) and, therefore, lessens the filtering required by the grating. With a 0.3-mm fused silica planar-planar etalon with R=96\% we have obtained a finesse of >50 with 90\% transmission albeit at 1300~nm. By reducing the etalon thickness and optimizing the reflection coating, the bandwidth of the broadband laser can be reduced and the FSR increased, improving robustness and efficiency for driving even weaker transitions. Moreover, the available output power is more than enough to properly seed a tapered amplifier, which can boost the output to the watt level.

Compared to DFB/DBR lasers, the presented solution offers several advantages. It is not limited to a very narrow wavelength range, it does not require a high-end current controller nor feedback from a wavemeter once it is set up, and it can be turned on and off on very short time scales, whereas a DFB/DBR laser will require several tens of seconds to settle once the current is switched on. For faster modulation, an external mechanical shutter (millisecond control) or an acousto-optic modulator would be called for, the latter being fast but having a limited extinction ratio. As for size, the BBL can also be made quite compact. An isolator is most often needed; in the BBL it is just put inside the cavity. The grating needs space, but it could be replaced by a small narrowband interference filter. Optical fiber can be coiled very tightly, and, of course, there is nothing preventing going very small with integrated optics. 

Compared to DFB/DBR lasers, external-cavity diode lasers (ECDLs) have the advantage of a reduced linewidth and broadband tunability. For driving dipole allowed transitions, the linewidth does not matter, and the tunability comes at the cost of reliability issues associated with mode hops and considerably added complexity in terms of cavity geometry, high-voltage piezo drivers and sophisticated control electronics. Mode hops resulting from fast on/off modulation would also be problematic. The BBL on the other hand, requires minimal electronics, and mode-hops are not an issue. Therefore, total power consumption can be very low especially in situations where continous operation is not required, something that is important in, e.g., space applications. 

Finally, strontium has a very low vapor pressure at room temperature, and at the high temperatures needed to obtain a sufficient vapor pressure for absorption spectroscopy, the Sr atoms attack glass surfaces, which prevents the use of simple vapor cells for laser stabilization. Frequently used alternatives such as dispenser sources \cite{Bridge2009a} and hollow cathode lamps \cite{Shimada2013a} are more complex and have a finite lifetime, which is a further motivation to seek frequency-stabilization-free solutions for applications requiring compact size and robustness.

\section{Conclusion}

In conclusion, a general design for broadband, frequency-stable lasers has been described that should find use especially in future transportable devices using trapped ions. While a ring laser with a long cavity operating in a multi-longitudinal-mode regime was presented, a linear and/or short cavity relying on a noisy, or modulated, driver for line broadening should also be an option. As long as the effective linewidth is much broader than the expected frequency drift remaining after passive frequency stabilization, no active frequency stabilization or sophisticated control electronics is needed, which, in turn, means that the device can be turned electronically on and off in a measurement cycle with sub-microsecond response times. The experimental fluorescence rates obtained for the Doppler-broadened 460.9-nm transition in Sr agree with theoretical expectations. The presented expressions are general and also valid for cold atoms and ions; they can thus be used to estimate the feasibility of replacing a narrowband laser with a broadband light source for any given transition.

\section*{Funding.} Academy of Finland (296476 and 306844).

\bibliography{IonClock}

\begin{thebibliography}{10}
\newcommand{\enquote}[1]{``#1''}

\bibitem{Haffner2008a}
H.~H\"affner, C.~Roos, and R.~Blatt, \enquote{Quantum computing with trapped
  ions,} {\protect\JournalTitle{Physics Reports}} \textbf{469}, 155 -- 203
  (2008).

\bibitem{Wineland2011a}
D.~J. Wineland and D.~Leibfried, \enquote{{Q}uantum information processing and
  metrology with trapped ions,} {\protect\JournalTitle{Laser Phys. Lett.}}
  \textbf{8}, 175 (2011).

\bibitem{March2009}
R.~E. March, \enquote{Quadrupole ion traps,} {\protect\JournalTitle{Mass
  Spectrometry Reviews}} \textbf{28}, 961--989 (2009).

\bibitem{Nolting2017}
D.~Nolting, R.~Malek, and A.~Makarov, \enquote{Ion traps in modern mass
  spectrometry,} {\protect\JournalTitle{Mass Spectrometry Reviews}} \textbf{0}
  (2017).

\bibitem{Margolis2009a}
H.~S. Margolis, \enquote{{T}rapped ion optical clocks,}
  {\protect\JournalTitle{Eur. Phys. J. Special Topics}} \textbf{172}, 97--107
  (2009).

\bibitem{Poli2013a}
N.~Poli, C.~W. Oates, P.~Gill, and G.~M. Tino, \enquote{{O}ptical atomic
  clocks,} {\protect\JournalTitle{Riv. Nuovo Cimento}} \textbf{36}, 555--624
  (2013).

\bibitem{Ludlow2015b}
A.~D. Ludlow, M.~M. Boyd, J.~Ye, E.~Peik, and P.~O. Schmidt, \enquote{{O}ptical
  atomic clocks,} {\protect\JournalTitle{Rev. Mod. Phys.}} \textbf{87},
  637--701 (2015).

\bibitem{Lindvall2013a}
T.~Lindvall, T.~Fordell, I.~Tittonen, and M.~Merimaa, \enquote{{U}npolarized,
  incoherent repumping light for prevention of dark states in a trapped and
  laser-cooled single ion,} {\protect\JournalTitle{Phys. Rev. A}} \textbf{87},
  013439 (2013).

\bibitem{Fordell2015a}
T.~Fordell, T.~Lindvall, P.~Dub\'{e}, A.~A. Madej, A.~E. Wallin, and
  M.~Merimaa, \enquote{{B}roadband, unpolarized repumping and clearout light
  sources for {S}r$^+$ single-ion clocks,} {\protect\JournalTitle{Opt. Lett.}}
  \textbf{40}, 1822--1825 (2015).

\bibitem{Kjaergaard2000a}
N.~Kjaergaard, L.~Hornekaer, A.~M. Thommesen, Z.~Videsen, and M.~Drewsen,
  \enquote{{I}sotope selective loading of an ion trap using resonance-enhanced
  two-photon ionization,} {\protect\JournalTitle{Appl. Phys. B}} \textbf{71},
  207 (2000).

\bibitem{Gulde2001a}
S.~Gulde, D.~Rotter, P.~Barton, F.~Schmidt-Kaler, R.~Blatt, and W.~Hogervorst,
  \enquote{{S}imple and efficient photo-ionization loading of ions for
  precision ion-trapping experiments,} {\protect\JournalTitle{Appl. Phys. B}}
  \textbf{73}, 861--863 (2001).

\bibitem{Johanning2011a}
M.~Johanning, A.~Braun, D.~Eiteneuer, C.~Paape, C.~Balzer, W.~Neuhauser, and
  C.~Wunderlich, \enquote{{R}esonance-enhanced isotope-selective
  photoionization of {Y}b{I} for ion trap loading,}
  {\protect\JournalTitle{Appl. Phys. B}} \textbf{103}, 327--338 (2011).

\bibitem{Lucas2004a}
D.~M. Lucas, A.~Ramos, J.~P. Home, M.~J. McDonnell, S.~Nakayama, J.-P. Stacey,
  S.~C. Webster, D.~N. Stacey, and A.~M. Steane, \enquote{{I}sotope-selective
  photoionization for calcium ion trapping,} {\protect\JournalTitle{Phys. Rev.
  A}} \textbf{69}, 012711 (2004).

\bibitem{Tanaka2005a}
U.~Tanaka, H.~Matsunishi, I.~Morita, and S.~Urabe, \enquote{{I}sotope-selective
  trapping of rare calcium ions using high-power incoherent light sources for
  the second step of photo-ionization,} {\protect\JournalTitle{Appl. Phys. B}}
  \textbf{81}, 795--799 (2005).

\bibitem{Wang2011b}
B.~Wang, J.~W. Zhang, C.~Gao, and L.~J. Wang, \enquote{{H}ighly efficient and
  isotope selective photo-ionization of barium atoms using diode laser and
  {LED} light,} {\protect\JournalTitle{Opt. Express}} \textbf{19}, 16438--16447
  (2011).

\bibitem{Bushaw2000a}
B.~Bushaw and W.~N\"ortersh\"auser, \enquote{{R}esonance ionization
  spectroscopy of stable strontium isotopes and $^{90}${S}r via
  $5\mathrm{s}^2\,^1\mathrm{S}_0 \rightarrow
  5\mathrm{s}5\mathrm{p}\,^1\mathrm{P}_1 \rightarrow
  5\mathrm{s}5\mathrm{d}\,^1\mathrm{D}_2 \rightarrow
  5\mathrm{s}11\mathrm{f}\,^1\mathrm{F}_3 \rightarrow \mathrm{Sr}^+$,}
  {\protect\JournalTitle{Spectrochim. Acta B}} \textbf{55}, 1679--1692 (2000).

\bibitem{Yasuda2006a}
M.~Yasuda, T.~Kishimoto, M.~Takamoto, and H.~Katori,
  \enquote{{P}hotoassociation spectroscopy of $^{88}\mathrm{Sr}$:
  {R}econstruction of the wave function near the last node,}
  {\protect\JournalTitle{Phys. Rev. A}} \textbf{73}, 011403 (2006).

\bibitem{Brownnutt2007a}
M.~Brownnutt, V.~Letchumanan, G.~Wilpers, R.~C. Thompson, P.~Gill, and A.~G.
  Sinclair, \enquote{{C}ontrolled photoionization loading of $^{88}${S}r$^+$
  for precision ion-trap experiments,} {\protect\JournalTitle{Appl. Phys. B}}
  \textbf{87}, 411--415 (2007).

\bibitem{Schuck2010a}
C.~Schuck, M.~Almendros, F.~Rohde, M.~Hennrich, and J.~Eschner,
  \enquote{{T}wo-color photoionization of calcium using {SHG} and {LED} light,}
  {\protect\JournalTitle{Appl. Phys. B}} \textbf{100}, 765--771 (2010).

\bibitem{Lawton2001a}
K.~Lawton and S.~Patterson, \enquote{{L}ong-term relative stability of
  thermistors,} {\protect\JournalTitle{Precision Engineering}} \textbf{25}, 24
  -- 28 (2001).

\bibitem{Lawton2002a}
K.~Lawton and S.~Patterson, \enquote{{L}ong-term relative stability of
  thermistors: {P}art 2,} {\protect\JournalTitle{Precision Engineering}}
  \textbf{26}, 340 -- 345 (2002).

\bibitem{Dinneen1999a}
T.~P. Dinneen, K.~R. Vogel, E.~Arimondo, J.~L. Hall, and A.~Gallagher,
  \enquote{{C}old collisions of $\mathrm{Sr}^*-\mathrm{Sr}$ in a
  magneto-optical trap,} {\protect\JournalTitle{Phys. Rev. A}} \textbf{59},
  1216--1222 (1999).

\bibitem{Lindvall2012a}
T.~Lindvall, M.~Merimaa, I.~Tittonen, and A.~A. Madej, \enquote{{D}ark-state
  suppression and optimization of laser cooling and fluorescence in a trapped
  alkaline-earth-metal single ion,} {\protect\JournalTitle{Phys. Rev. A}}
  \textbf{86}, 033403 (2012).

\bibitem{Blushs2004a}
K.~Blushs and M.~Auzinsh, \enquote{{V}alidity of rate equations for {Z}eeman
  coherences for analysis of nonlinear interaction of atoms with broadband
  laser radiation,} {\protect\JournalTitle{Phys. Rev. A}} \textbf{69}, 063806
  (2004).

\bibitem{Safronova2013a}
M.~S. Safronova, S.~G. Porsev, U.~I. Safronova, M.~G. Kozlov, and C.~W. Clark,
  \enquote{{B}lackbody-radiation shift in the {S}r optical atomic clock,}
  {\protect\JournalTitle{Phys. Rev. A}} \textbf{87}, 012509 (2013).

\bibitem{Bridge2009a}
E.~M. Bridge, J.~Millen, C.~S. Adams, and M.~P.~A. Jones, \enquote{{A} vapor
  cell based on dispensers for laser spectroscopy,} {\protect\JournalTitle{Rev.
  Sci. Instrum.}} \textbf{80}, 013101 (2009).

\bibitem{Shimada2013a}
Y.~Shimada, Y.~Chida, N.~Ohtsubo, T.~Aoki, M.~Takeuchi, T.~Kuga, and Y.~Torii,
  \enquote{{A} simplified 461-nm laser system using blue laser diodes and a
  hollow cathode lamp for laser cooling of {S}r,} {\protect\JournalTitle{Rev.
  Sci. Instrum.}} \textbf{84}, 063101 (2013).

\end{thebibliography}






\end{document}